\documentclass[%
 aip,
rsi,%
 amsmath,amssymb,
reprint,%
]{revtex4-1}
\usepackage{color}
\usepackage{graphicx}
\usepackage{graphics}
\usepackage{epsfig}
\usepackage{amsmath}
\usepackage{array}
\usepackage{amssymb}
\usepackage{graphicx}      
\usepackage{dcolumn}       
\usepackage{bm}            
\usepackage{array}
\usepackage{booktabs}
\usepackage{ulem}
\usepackage{braket}

\usepackage[linktocpage=true,
  colorlinks=true, 
  pdfborder={0 0 0},
  linkcolor=blue,
  citecolor=red,
  filecolor=yellow,
  urlcolor=blue,
  bookmarks,
  pdfauthor={},
]{hyperref}

\usepackage[dvipsnames]{xcolor}
\newcommand{\commentyf}[1]{\textcolor{ForestGreen}{{#1}}}

\newcommand{\tokyo}{Department of Applied Physics, The University of Tokyo, Tokyo 113-8656, Japan}

\begin{document}

\title{Quasiparticle energy spectra of isolated atoms from coupled-cluster singles and doubles (CCSD): Comparison with exact CI calculations}

\author{Hirofumi Nishi}   \affiliation{\tokyo}
\author{Taichi Kosugi}   \affiliation{\tokyo}
\author{Yoritaka Furukawa} \affiliation{\tokyo} 
\author{Yu-ichiro Matsushita}   \email{matsushita@ap.u-tokyo.ac.jp} \affiliation{\tokyo}

\date{\today}

\begin{abstract}
In this study, we have calculated single-electron energy spectra via the Green's function based on the coupled-cluster singles and doubles (GFCCSD) method for isolated atoms from H to Ne. In order to check the accuracy of the GFCCSD method, we compared the results with the exact ones calculated from the full-configuration interaction (FCI). Consequently, we have found that the GFCCSD method reproduces not only the correct quasiparticle peaks but also satellite ones by comparing the exact spectra with the 6-31G basis set. It is also found that open-shell atoms such as C atom exhibit Mott gaps at the Fermi level, which the exact density-functional theory (DFT) fails to describe. The GFCCSD successfully reproduces the Mott HOMO-LUMO (highest-occupied molecular orbital and lowest-unoccupied molecular orbital) gaps even quantitatively. We also discussed the origin of satellite peaks as shake-up effects by checking the components of wave function of the satellite peaks. The GFCCSD is a novel cutting edge to investigate the electronic states in detail.
\end{abstract}

\pacs{Valid PACS appear here}
\maketitle

\section{Introduction}
An electron in materials behaves as a quasiparticle occupying a discrete energy level, called single-electron spectra or quasiparticle energy spectra, following the quantum mechanics. Single-electron spectra are one of the most fundamental quantities and a prerequisite for understanding the electronic properties of the material. Each energy peak in single-electron spectra is a consequence of the chemical bonds in the material and teaches us the fine information of the material. In fact, the understanding of single-electron spectra is the first step for further analysis of the material properties. Single-electron spectrum is classified into two categories: quasiparticle peak and satellite peak. Importantly, single-electron spectrum is experimentally observed by X-ray photoelectron spectroscopy (XPS) \cite{ARPES}.

From the theoretical viewpoint, the density-functional theory (DFT) is one of the most successful tools to draw single-electron spectrum in finite systems and solids \cite{HK,KS}. In spite of the relatively cheap calculation cost, the DFT provides proper structural and electronic properties of the material. However, the DFT cannot reproduce energy gaps in strongly correlated systems, van der Waals interactions, and satellite peaks because of the mean-field nature. Recently, many efforts have been done to solve the difficulties: self-interaction-error correction (SIC) method \cite{SICDFT}, LDA+U \cite{LDAU}, hybrid functionals \cite{hybrid1,hybrid2,hybrid3} , LDA+DMFT \cite{DMFT1,DMFT2,DMFT3}, GW \cite{GW1,GW2,GW3}, GW+cumulant expansion \cite{GWC1,GWC2}, van der Waals DFT \cite{DFTvdW1,DFTvdW2}, RDMFT \cite{RDMFT1,RDMFT2,Sangeeta}, etc. Still, the development of a novel methodology for highly accurate self-consistent electronic-structure calculations stays a central and important problem in theoretical material science.

On the other hand, wave function theory (WFT) is frequently used for (mainly) finite systems in the quantum chemistry field. Compared with the DFT, the WFT has a great advantage of high accuracy and we can easily improve the accuracy. With these advantages, recently, WFT is getting more attention also in the field of physics such as density-matrix-renormalization group (DMRG) \cite{DMRG1,DMRG2}, transcorrelated method\cite{TC1,TC5,TC6}, and Monte-Carlo configuration-interaction approach \cite{FCIQMC,NiOCCSD}. In the WFT, full-configuration interaction (FCI) approach is the most important method yielding the exact total energy and the wave function. However, FCI calculations require huge computational cost, and it cannot be applied to larger systems.
Another notable WFT is coupled-cluster (CC) theory \cite{CCSD1,CCSD2,CCSD3}. It efficiently involves many Slater determinants expanded as Taylor series of an excitation operator, which results in a smaller computational cost than FCI (see detailed explanations in Sec.~\ref{method}). Especially, it is reported that the CC singles and doubles (CCSD), which introduces an approximation to truncate the excitation operators up to double excitation levels in CC theory can reproduce the spin gap of NiO known as a strongly-correlated periodic system \cite{NiOCCSD}. It is a notable thing that most many-body WFT cannot describe the single-electron spectra directly (with a few exceptions such as transcorrelated method) and that most previous studies with WFT focus only on the ground state energy. Electronic excited states can also be calculate in CC theory by using the equation-of-motion CC (EOM-CC) \cite{EOM1,EOM2} or symmetry-adapted cluster/configuration interaction (SAC-CI) \cite{SACCIPk} method. However, in most cases, only the 1st ionization potential or the 1st electron affinity get interests, and the number of the studies on single-electron spectra within the CC theory is quite small \cite{UEGCCGF}.

One-particle Green's function is known to be useful for grasping the electronic properties. It is because the one-particle Green's function is enough to calculate all the one-body physical quantities including electron density and even total energy. Furthermore, from the one-particle Green's function, we can calculate the single-electron energy spectra. The formalism in which the one-particle Green's function is constructed from the CC theory (GFCC) was proposed \cite{Nooijen92,Nooijen93,Nooijen95}, and was applied to a uniform electron gas \cite{UEGCCGF}. However, the number of the works reporting actual application is quite few. In addition, the accuracy of the GFCC has not been studied yet. Therefore, comparative study for understanding the GFCC itself comparing with the exact spectra is necessary.

Our motivation of this work is to clarify the accuracy of single-electron energy spectra from the GFCCSD in comparison with those of the FCI calculations for isolated atoms from H to Ne. In the calculated single-electron spectra, we also found several satellite peaks. Then, we discussed the electronic structure of the satellite peaks and clarified the origin of the satellite peaks. Furthermore application of this method is extended to $d$-elements and periodic materials will be reported in other works\cite{Kosugi,Furukawa}.

The organization of this paper is as follows. We explain the summary of the GFCCSD method in section \ref{method}. In subsection \ref{result1}, we show the accuracy of the GFCCSD method by comparing with the FCI calculation. The origin of the satellite peaks is clarified in section \ref{result2}. We performed comparative systems using more accurate basis set (cc-pVDZ) and compared with the obtained experimental results in section \ref{result3}. Section \ref{conclusion} summarizes our findings.

\section{method}\label{method}
In subsection~\ref{CCGF}, we briefly describe the GFCC method introduced by Nooijen \cite{Nooijen92,Nooijen93,Nooijen95}. The following subsection \ref{CIGF} shows the recipe of the construction of the Green's function from the CI method.
\subsection{Green's function from the coupled-cluster calculations}\label{CCGF}
Throughout this study, we focus only on the non-relativistic Hamiltonian, $\hat{H}$. 
The wave function in the coupled-cluster theory is expressed as  $\ket{\Psi_{\rm CC}} = \mathrm{e}^{\hat{T}}\ket{\Phi} \ , $
where the operator $\hat{T}$ represents the electron excitation written as
$\hat{T} 
= \sum_{ia} t_i^a \hat{a}_a^{\dagger} \hat{a}_i  
+ \frac{1}{4} \sum_{ijab} t_{ij}^{ab} \hat{a}_a^{\dagger} \hat{a}_b^{\dagger} \hat{a}_j \hat{a}_i  
+\cdots$.
The operator $\hat{a}_{p}$ is an annihilation operator and the $\hat{a}_p^{\dagger}$ is a creation operator. The index $i,j,\cdots$ represents occupied states, the index $a,b,\cdots$ represents unoccupied states, and the label $p,q,\dots$ are any states irrespective of occupied and unoccupied. The coefficients $t_{ij\cdots}^{ab\cdots}$ of the excitation  operators can be determined  from amplitude equations which are derived by projecting excited states $\bra{\Phi_{ij\cdots}^{ab\cdots}}$ to the Schr\"odinger equation, in which a similarity transformed Hamiltonian appears,
\begin{equation}
\braket{\Phi_{ij\cdots}^{ab\cdots}|\mathrm{e}^{-\hat{T}}\hat{H}\mathrm{e}^{\hat{T}}|\Phi}=0 \ .
\end{equation}
After determining the coefficients in ${\hat T}$, the total energy can be calculated by projecting $\bra{\Phi}$:
\begin{equation}
\braket{\Phi|\mathrm{e}^{-\hat{T}}\hat{H}\mathrm{e}^{\hat{T}}|\Phi}=E \ .
\end{equation}

One-particle Green's function of the frequency representation at zero temperature is written as
\begin{equation}
\begin{split}
  G_{pq}(\omega)
  &= G_{pq}^{(h)}(\omega) + G_{pq}^{(e)}(\omega) \\
  &=
   \braket{\Psi|\hat{a}_q^{\dagger}\frac{1}{\omega+\hat{H}_N}\hat{a}_p|\Psi}
   +
  \braket{\Psi|\hat{a}_q\frac{1}{\omega-\hat{H}_N}\hat{a}_p^{\dagger}|\Psi} 
   \ ,
\end{split}
\label{def_green}
\end{equation}
where the Green's function can be separated into the electron removal and attachment part (partial Green's functions). The $\hat{H}_N$ is defined as $\hat{H}_N=\hat{H}-E_0$, where $E_0$ is the total energy of the exact ground state described as $\ket{\Psi}$. Here, one adopts the CCSD wave function to the exact wave function, $\ket{\Psi}=\ket{\Psi_{\rm CC}}$.
Using the similarity transformed Hamiltonian 
$\bar{H}_{N}
=\mathrm{e}^{-\hat{T}}\hat{H}\mathrm{e}^{\hat{T}}-E_0$ and the transformed creation and annihilation operators $\bar{a}_q^{\dagger}=\mathrm{e}^{-\hat{T}} \hat{a}_q^{\dagger}\mathrm{e}^{\hat{T}}$ and $\bar{a}_p
=\mathrm{e}^{-\hat{T}} \hat{a}_p\mathrm{e}^{\hat{T}}$, we can rewrite the partial Green's functions to
\begin{equation}
\label{partial_hole}
G_{pq}^{(h)}(\omega)
= \braket{\Phi|(1+\hat{\Lambda})\bar{a}_p^{\dagger}
\frac{1}{\omega+\bar{H}_N}
\bar{a}_q|\Phi} ,
\end{equation}
\begin{equation}
\label{partial_ele}
G_{pq}^{(e)}(\omega)
= \braket{\Phi|(1+\hat{\Lambda})\bar{a}_p
\frac{1}{\omega-\bar{H}_N}
\bar{a}_q^{\dagger}|\Phi} .
\end{equation}
Note that the transformed Hamiltonian $\bar{H}_{N}$ is not Hermitian and that the GFCC is constructed using bi-variational method \cite{Arponen83,Bi-vari1,Bi-vari2}. The operator $\hat{\Lambda}$ is a de-excitation operator which is determined by
\begin{equation}
\braket{\Phi_{ij\cdots}^{ab\cdots}|(1+\hat{\Lambda})
\mathrm{e}^{-\hat{T}}{\hat{H}}\mathrm{e}^{\hat{T}}|\Phi}=0 \ .
\end{equation}

In order to avoid the computational difficulty in treating the inverse matrix $(\omega \pm {\bar H}_N)^{-1}$ of Eqs.~(\ref{partial_hole}) and (\ref{partial_ele}), $\hat{X}_q(\omega)$ and $\hat{Y}_q(\omega)$ are introduced as follows
\begin{equation}
(\omega+\bar{H}_N)\hat{X}_q(\omega)\ket{\Phi} 
= \bar{a}_q\ket{\Phi},
\label{eq:ipgf}
\end{equation}
\begin{equation}
(\omega-\bar{H}_N)\hat{Y}_q(\omega)\ket{\Phi} 
= \bar{a}_q^{\dagger}\ket{\Phi}.
\label{eq:eagf}
\end{equation}
Once we solve Eq.~(\ref{eq:ipgf}) and (\ref{eq:eagf}), we can get the information of the involving ($N-1$)- and ($N+1$)-electron states, respectively. 
Note that these two linear equations are equivalent to Hamiltonian of EOM-CC theory: Eq.~(\ref{eq:ipgf}) corresponds to ($N-1$)-electron states yielding ionization potential (IP-EOM-CC) and Eq.~(\ref{eq:eagf}) corresponds to ($N+1$)-electron states (EA-EOM-CC).
With $\hat{X}_q(\omega)$ and $\hat{Y}_q(\omega)$, the  Green's function is finally expressed as \cite{Kowalski14,Kowalski16} 
\begin{equation}
  G_{pq}^{(h)}(\omega)
  = \braket{\Phi|(1+\hat{\Lambda})\bar{a}_p^{\dagger}\hat{X}_{q}(\omega)|\Phi},
\end{equation}
\begin{equation}
  G_{pq}^{(e)}(\omega)
  = \braket{\Phi|(1+\hat{\Lambda})\bar{a}_p \hat{Y}_{q}(\omega)|\Phi}.
\end{equation}
We can calculate single-electron spectra by virtue of the Green's function:
\begin{equation}
A(\omega)=-\frac{1}{\pi}{\rm Im}\left[{\rm tr} \left({G}(\omega+i\delta)\right)\right].
\label{spectr}
\end{equation}

In this study, we truncate the excitation operator ${\hat T}$ up to singles and doubles (CCSD) as follows:
\begin{equation}
 \hat{T} 
\simeq \sum_{ia}t_i^a\hat{a}_a^{\dagger}\hat{a}_i  
+ \frac{1}{4}\sum_{ijab}t_{ij}^{ab}\hat{a}_a^{\dagger}\hat{a}_b^{\dagger}\hat{a}_j\hat{a}_i.
\end{equation}
By introducing the truncation in the ${\hat T}$ operator, we derive the following equations for ${\hat \Lambda}$, ${\hat X}_q$, and ${\hat Y}_q$ operators maintaining the same accuracy as CCSD:
\begin{equation}
 \hat{\Lambda} 
\simeq  \sum_{ia}\lambda_i^a\hat{a}_i^{\dagger}\hat{a}_a  
+ \frac{1}{4}\sum_{ijab}\lambda_{ij}^{ab}\hat{a}_i^{\dagger}\hat{a}_j^{\dagger}\hat{a}_b\hat{a}_a  
\end{equation}
\begin{equation}
 \hat{X}_q(\omega) 
\simeq  \sum_{i}x_{i(q)}(\omega)\hat{a}_i  
+ \frac{1}{2}\sum_{ija}x_{ij(q)}^{a}(\omega)\hat{a}_a^{\dagger}\hat{a}_j\hat{a}_i  
\end{equation}
\begin{equation}
 \hat{Y}_q(\omega) 
\simeq  \sum_{a}y_{a(q)}(\omega)\hat{a}_a^{\dagger}  
+ \frac{1}{2}\sum_{iab}y_{i(q)}^{ab}(\omega)\hat{a}_a^{\dagger}\hat{a}_b^{\dagger}\hat{a}_i  .
\end{equation}
In particular, ${\hat X}_q$ operators are truncated up to $1h$ (1st term of the right hand side (r.h.s.)) and $2h1p$ term (2nd term of the r.h.s.), and ${\hat Y}_q$ operators are similarly truncated up to $1p$ (1st term of the r.h.s.) and $2p1h$ term (2nd term of the r.h.s.). These truncation for ${\hat X}_q$ and ${\hat Y}_q$ leads to the expression of the wave function after electron attachment/removal to be
\begin{eqnarray}
| \Psi^{N-1} \rangle &=&e^{\hat T}\sum_i x_{i(q)}(\omega){\hat a}_i\left|\Phi\right>+e^{\hat T}\sum_{ija} x_{ij(q)}^a(\omega){\hat a}^\dagger_a{\hat a}_j{\hat a}_i\left|\Phi\right> \nonumber\\
&\equiv&e^{\hat T}\sum_{1h}\left|1h\right>+e^{\hat T}\sum_{2h1p}\left|2h1p\right>
\label{def_Psi_N-1}
\end{eqnarray}
\begin{eqnarray}
| \Psi^{N+1} \rangle &=&e^{\hat T}\sum_a y_{a(q)}(\omega){\hat a}_a^\dagger\left|\Phi\right>+e^{\hat T}\sum_{iab} y_{i(q)}^{ab}(\omega){\hat a}^\dagger_a{\hat a}^\dagger_b{\hat a}_i\left|\Phi\right> \nonumber\\
&\equiv&e^{\hat T}\sum_{1p}\left|1p\right>+e^{\hat T}\sum_{2p1h}\left|2p1h\right>,
\label{def_Psi_N+1}
\end{eqnarray}
where we introduced notations describing subspace in Hilbert space, $\left|1h\right>$, $\left|2h1p\right>$, $\left|1p\right>$, and $\left|2p1h\right>$, representing 1 electron annihilated, 1 electron annihilated and 1 electron excited, 1 electron created, and 1 electron created and 1 electron excited from the HF electron configuration, respectively. 

The computational cost is $O(N^6)$ for the CCSD and $\Lambda$-CCSD calculations. Solving the IP/EA-EOM-CCSD linear equations is computationally demanding. We solve these equations using the shifted bi-conjugate gradient (Bi-CG) method \cite{Frommer03} which takes $O(N^6)$. However, in case of the shifted Bi-CG method not converged, we use LU-decomposition costing $O(N^9N_{\omega})$ where $N_{\omega}$ is the number of $\omega$ mesh.

\subsection{Green's function from the FCI calculations}\label{CIGF}

For comparison, we have also calculated the exact single-electron spectra from the FCI calculations. We explain the calculation procedure for the construction of the one-body Green's function (GF) briefly below by using the many-body ground state(s) obtained via exact diagonalization\cite{ARPACK}.

The exact expression for the GF at zero temperature is written as\cite{PhysRevB.76.245116} 
\begin{gather}
	G_{p q} (\omega)
	=
		\frac{1}{Z}
		\sum_\nu
			[
				G_{p q}^{(e) \nu} (\omega)
				+
				G_{p q}^{(h) \nu} (\omega)
			]
            ,
	\label{G_imag_freq_sum_partial_G}
\end{gather}
corresponding to Eq. (\ref{def_green}) for the GFCC,
where 
\begin{gather}
	G_{p q}^{(e) \nu} (\omega)
	=
		\langle \Psi^{\mathrm{FCI}}_\nu | \hat{a}_p 
		\frac{1}{ \omega + \varepsilon_\nu - \hat{H}}
		\hat{a}_q^\dagger | \Psi^{\mathrm{FCI}}_\nu \rangle
	\label{def_partial_G_e}
\end{gather}
and 
\begin{gather}
	G_{p q}^{(h) \nu} (\omega)
	=
		\langle \Psi^{\mathrm{FCI}}_\nu | \hat{a}_q^\dagger 
		\frac{1}{ \omega - \varepsilon_\nu + \hat{H}}
		\hat{a}_p | \Psi^{\mathrm{FCI}}_\nu \rangle
	\label{def_partial_G_h}
\end{gather}
are the partial GFs for electron and hole excitations, respectively,
from the $\nu$th many-body lowest-energy state.
The summation on the right-hand side in Eq.~(\ref{G_imag_freq_sum_partial_G}) is for the lowest-energy states and $Z$ is the partition function.

The individual components of the GF can be calculated by employing the ordinary Lanczos method\cite{Lanczos_for_exact_diag}, which is applicable originally only to the diagonal components. Specifically, we calculate the following auxiliary quantities:
\begin{gather}
	G_{p q}^{\pm (e) \nu} (\omega)
	\equiv
		\langle \Psi^{\mathrm{FCI}}_\nu | \hat{a}_{p q}^\pm
		\frac{1}{ \omega + \varepsilon_\nu - \hat{H}}
		\hat{a}_{p q}^{\pm \dagger} | \Psi^{\mathrm{FCI}}_\nu \rangle
      ,
\end{gather}
where
\begin{gather}
	\hat{a}_{p q}^+
    \equiv
	    \hat{a}_p + (1 + i) \hat{a}_q
       \\
	\hat{a}_{p q}^-
    \equiv
		\hat{a}_p + (1 - i) \hat{a}_q
        .
\end{gather}
From these auxiliary quantities,
we can obtain not only the contributions from the electron excitations to the diagonal components but also those to the off-diagonal components in the GF as 
\begin{gather}
	G_{p q}^{(e) \nu} (\omega)
	=
		\frac{1 + i}{4}
			G_{p q}^{+ (e) \nu} (\omega)
		+
		\frac{1 - i}{4}
			G_{p q}^{- (e) \nu} (\omega)
           	\nonumber \\
		-
		\frac{1}{2}
			G_{p p}^{(e) \nu} (\omega)
		-
			G_{q q}^{(e) \nu} (\omega)
	\label{partial_G_e_ij}
	\\
	G_{q p}^{(e) \nu} (\omega)
	=
		\frac{1 - i}{4}
			G_{p q}^{+ (e) \nu} (\omega)
		+
		\frac{1 + i}{4}
			G_{p q}^{- (e) \nu} (\omega)
            \nonumber \\
		-
		\frac{1}{2}
			G_{p p}^{(e) \nu} (\omega)
		-
			G_{q q}^{(e) \nu} (\omega)
	\label{partial_G_e_ji}
    .
\end{gather}
The contributions from the hole excitations can also be calculated similarly.

\section{results}\label{result}
We describe the detailed analysis of the accuracy of the GFCC in subsection~\ref{result1} comparing with FCI calculations. Subsection~\ref{result2} gives the detailed analzyses for the origins of the satellite peaks. In subsection~\ref{result3}, Single-electron spectra from H to Ne atom are presented.
\subsection{Comparison with FCI}\label{result1}
\begin{figure}
\includegraphics[width=1.0\linewidth]{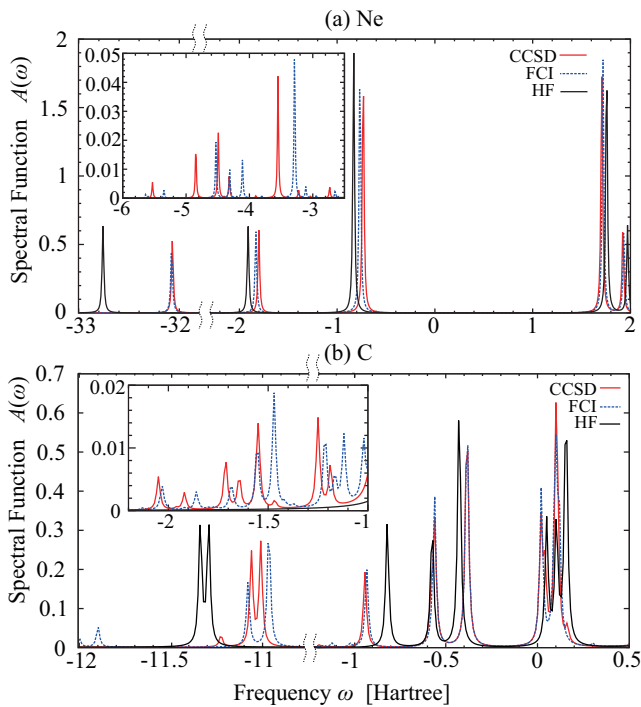}
\caption{(Color online) Single-particle spectra of Ne (a) and C (b) atoms calculated based on the GFCCSD (red line), the FCI (blue dotted line), and the HF (black line) with the 6-31G basis set. The insets show the enlarged satellite peaks. The vacuum level is set to 0. The energy levels of the highest-occupied molecular orbital (HOMO) for Ne atom are $-0.73$ Hartree for GFCCSD, $-0.77$ Hartree for FCI, and $-0.792$ Hartree for the experimental value\cite{experiment}, respectively, and the HOMO levels for C atom are $-0.38$ Hartree for GFCCSD and FCI and $-0.414$ Hartree for experimental value \cite{experiment}, respectively.}
\label{Fig1}
\end{figure}

First, we have checked the accuracy of the GFCCSD by comparing with those of the FCI. Since the computational cost of the FCI is quite huge, we adopted the 6-31G basis set including 1$s$, 2$s$, 2$p$, $3s$, and $3p$ orbitals for each element in this subsection. As an example, we show the results of isolated C and Ne atoms in Fig.~\ref{Fig1}. The reference wave functions $\ket{\Phi}$ of the Ne and C atoms were calculated by the unrestricted Hartree--Fock (UHF) method.

\begin{figure}
\includegraphics[width=1.0\linewidth]{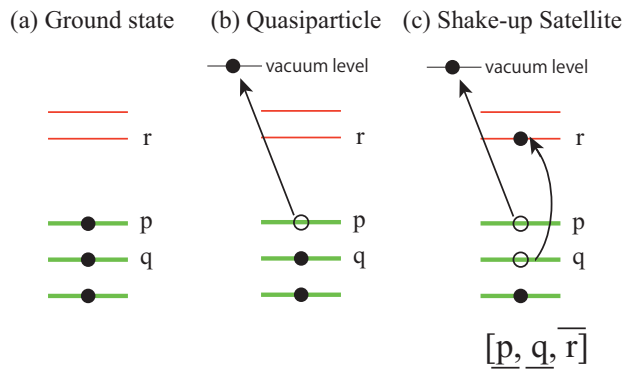}
\caption{(Color online) Schematic picture of electron configuration of (a) ground state, (b) quasiparticle, \commentyf{and} (c) Shake-up satellite. Green line represent an occupied orbital (including spin degrees of freedom) and red line is unoccupied orbitals in the ground state. Black filled circle denotes an electron and black circle is a hole. In particular, figure (b) and (c) show the main contribution of quasiparticle peak and of shake-up satellite one, respectively. In (c), the shake-up satellite denoted by a notation
[$\underline{p},\underline{q},\overline{r}$] is shown (see the text).}
\label{Fig_Satellite}
\end{figure}

We can see clear and sharp quasiparticle peaks. The spectra of Ne atom have five major peaks corresponding to 1$s$, 2$s$, 2$p$, 3$p$, and 3$s$ in ascending order of the frequencies. This identification of the character of each quasiparticle peak was confirmed by checking the wave function character at the corresponding energy. As stated in Sec.~\ref{method}, the wave function of the ($N-1$)- or ($N+1$)-electron system is expressed as the linear combination of $e^{\hat T}\left|1h\right>$, $e^{\hat T}\left|2h1p\right>$, $e^{\hat T}\left|1p\right>$, and $e^{\hat T}\left|2p1h\right>$ parts [see Eqs.~(\ref{def_Psi_N-1}) and (\ref{def_Psi_N+1})]. By looking at these components of the wave function, we can identify the main contributions to the quasiparticle peaks. We found that the quasiparticle peaks are attributed to $e^{\hat T}\ket{1h}$ or $e^{\hat T}\ket{1p}$, in which the removed/attached electron orbital corresponds to each quasiparticle peak (see schematic picture of Fig.~\ref{Fig_Satellite}(b)).  
The differences in the intensities of the quasiparticle peaks manifest the degrees of degeneracy. We can see the large peaks at $-0.77$ Hartree of triply degenerate 2$p$ orbitals in Fig.~\ref{Fig1} (a). The quasiparticle peaks of Ne atom are in good agreement with those of the FCI in Fig.~\ref{Fig1} (a): The HOMO-LUMO (highest-occupied molecular orbital and lowest-unoccupied molecular orbital) gap is 2.44 Hartree for the GFCCSD method and 2.49 Hartree for the FCI method, which are narrower than that of the Hartree--Fock (HF), 2.59. This HOMO-LUMO gap narrowing is consistent with the well-known fact that the HF method tends to overestimate energy gaps in general. Moreover, we can also see satellite peaks in GFCCSD [see the insets of Fig.~\ref{Fig1}(a)]. The peak positions in energy of GFCCSD deviate from those of FCI. However, it is found that the GFCCSD reproduces the correct number of satellite peaks and their rough positions. The detailed analysis of the satellite peaks is given in subsection~\ref{result2}.

\begin{figure}
\includegraphics[width=1.0\linewidth]{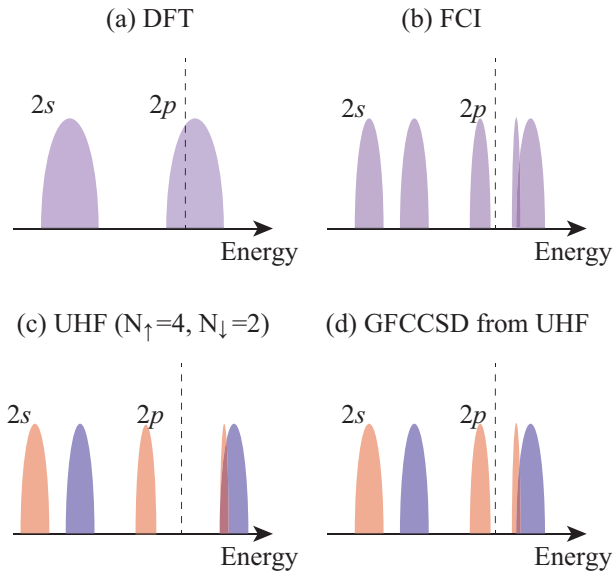}
\caption{(Color online) Schematic picture of density of states (DOS) for C atom by (a) DFT, (b) FCI, (c) UHF, and (c) GFCCSD from the UHF. DOS colored by purple represents the spin-degeneracy. Blue (orange) DOS depicts up- (down-)spin DOS. In (c), 2$s$ and 2$p$ orbitals are splitted into two peaks breaking the spin symmetry. Dotted line in each figure represents the Fermi energy.}
\label{Fig_DOS}
\end{figure}

In Fig.~\ref{Fig1}(b), the quasiparticle peaks of C atom by FCI and GFCCSD are in good agreement, and the overestimation of the HOMO-LUMO gap by HF is improved in the FCI and the GFCCSD: $0.40$ Hartree for the GFCCSD method, $0.40$ Hartree for the FCI method, and $0.48$ Hartree for the HF method. Interestingly, each quasiparticle peak of 1$s$ and 2$s$ splits into two peaks in contrast to the Ne case. The sharp two peaks around $-11$ Hartree correspond to the 1$s$ orbital and those from $-1$ to $-0.5$ Hartree correspond to the 2$s$ orbital. In the case of HF, we found that the energy splitting is the consequence of the breaking of the spin symmetry, namely exchange splitting. By considering the Hund's rule, the C atom is in a spin-triplet state for the ground state. Since we do not incorporate the relativistic effects for our calculations, the three triplet states, $S_z=-1,0,1$, are degenerate. In the UHF case, such a spin symmetry is broken and an arbitrary one state is chosen spontaneously, leading to the exchange splitting for each of the $1s$ and $2s$ orbitals (see the schematic picture in Fig.~\ref{Fig_DOS}(c). On the other hand, the FCI method obtained nine-fold degenerate ground states and the spin symmetry is preserved as a whole. Nevertheless, energy splitting in $1s$ and $2s$ orbitals are observed also in the FCI interestingly. Focusing on the results of GFCCSD, the calculated energy spectra reproduce the FCI results with high accuracy. In contrast, these energy splittings cannot appear in the exact or symmetry-conserved DFT calculation (see the schematic picture in Fig.~\ref{Fig_DOS}(a)). We performed the DFT calculation with real-space grids along the radial direction using the Tokyo {\it Ab initio} Program Package (TAPP) \cite{TAPP1,TAPP2,TAPP3}. We employed PBE functional \cite{PBE96} for the exchange-correlation energy. The TAPP code treats a fractional occupation to recover the exact spin and spherical symmetry. Thus, $p_x$, $p_y$, and $p_z$ orbitals are occupied by 2/3 electrons in the C atom. The calculated single-electron energy level is $-10.071$ Hartree for 1$s$, $-0.509$ Hartree for 2$s$, $-0.197$ Hartree for 2$p$ with triple degeneracy, thus showing neither energy splitting in 1$s$ and 2$s$ orbitals.

Another interesting point is the appearance of the HOMO-LUMO gap in C atom. C atom is an example of open shell systems. With the symmetry-conserved DFT, the Fermi level  crosses the 2$p$ orbitals as shown in Fig.~\ref{Fig_DOS}(a): Two electrons partially fill the $p_x$, $p_y$, and $p_z$ orbitals. In contrast, FCI calculations show the HOMO-LUMO gap at the Fermi level (Fig.~\ref{Fig_DOS}(b)). It is noteworthy that FCI causes the energy gap without breaking the spherical nor spin symmetry: Mott gap. The UHF calculation also exhibit the HOMO-LUMO gap. We have found that it is derived from the breaking of spherical symmetry. In UHF calculation, two electrons occupy two of the three $2p$ orbitals leading to the breaking of spherical symmetry (Fig.~\ref{Fig_DOS}(c)). We have found that the GFCCSD method succeed to the property of the UHF. However, interestingly, the GFCCSD reproduces the Mott gap spectrum quite well. If we use the GFCCSD method combined with the multi-reference CCSD method, then it is expected that the GFCCSD method might be free from the limitation of UHF and show the Mott gap without breaking the spherical symmetry.

We also see several small peaks at sides of each tall quasiparticle peaks called satellite peaks (resonance states) whose intensities are smaller than $0.1$. Satellite peaks around near the 2$s$ peaks are shown in the insets of Fig.~\ref{Fig1}(a) and (b).
In the presence of the satellite peaks, the intensity weight of the quasiparticle peaks becomes smaller than 1. The intensity weight of the satellite peaks is known as a renormalization factor indicating how the material is strongly correlated \cite{AGD}. 
\begin{equation}
  N_{\rm satellite}=\int_{\rm satellite} d\omega A(\omega) \ .
\end{equation}
The renormalization factor, $N_{\rm satellite}$, calculated from the GFCCSD of the C atom is $0.72$ and that of the Ne atom is $0.65$. In the C atom, overall features (position and intensity) of the satellite peaks in the GFCCSD are confirmed with the peaks in the FCI. However, the positions of the satellite peaks are shifted from those of the CCSDGF. The detail of these satellite peaks is discussed in the next subsection.

\subsection{The origin of satellite peaks}\label{result2}

\begin{table}[htb]
\renewcommand{\arraystretch}{1.3}
  \caption{Satellite-peak position in the unit of Hartree for Ne obtained with the GFCCSD, FCI, and estimated by the HF (see the text).}
  \begin{tabular}{lcccc} \hline \hline
    Main configuration~~~~~~~~~~~~  &  ~CCSDGF~ &~~~FCI~~~~& ~~~HF~~~~\\ \hline 
  
$[\underline{2p}$,$\underline{2p}$,$\overline{3p}]$ & -3.55 & -3.29 & -2.92 \\
$[\underline{2s}$,$\underline{2p}$,$\overline{3s}]$ & -4.32 & -4.11 & -4.16 \\
$[\underline{2s}$,$\underline{2p}$,$\overline{3s}]$ & -4.39 & -4.31 & -4.24 \\
$[\underline{2s}$,$\underline{2p}$,$\overline{3p}]$ & -4.85 & -4.53 & -4.05 \\ 
$[\underline{2s}$,$\underline{2s}$,$\overline{3s}]$ & -5.33 & -5.35 & -5.33 \\ \hline\hline
\end{tabular}
\label{Table1}
\end{table}

Table~\ref{Table1} shows the positions of the satellite peaks in the valence-band side of Ne atom only with higher intensity over 0.03. The peak positions calculated from the GFCCSD and FCI are presented in the Table. In the same way as the analysis in subsec.~\ref{result1}, by looking at the CCSD wave function, we can identify the main contribution to the satellite peaks. Then, we have found that the wave function of the satellite peaks is mainly attributed to the $e^{\hat T}\ket{2h1p}$ and $e^{\hat T}\ket{2p1h}$ parts. The $e^{\hat T}\ket{2h1p}$ part is the main contributor to the satellite peaks in the occupied-level side, while the $e^{\hat T}\ket{2p1h}$ contributes to the satellite peaks in the unoccupied-level side. $e^{\hat T}\ket{2h1p}$ terms represent 1 electron annihilation associated with 1 electron excitation known as shake-up satellite (see schematic picture in Fig.~\ref{Fig_Satellite}(c)). The energy position of the shake-up satellites deviates from the quasiparticle peaks by the excitation of an electron. We labeled each satellite peaks by three orbitals corresponding to $\ket{2h1p}$: 2 occupied (hole) and 1 unoccupied (particle) orbitals contributing to the satellite peak the most. We introduced an annotation here to describe the main contribution to $\ket{2h1p}$. A combination of three orbitals, [$\underline{p},\underline{q},\overline{r}$], describes that one electron occupying the orbital $p$ is removed from the system in association with another electron occupying $q$ orbital being excited to the orbital $r$, in which the underlines (overline) indicate occupied (unoccupied) orbitals explicitly (see schematic picture in Fig.~\ref{Fig_Satellite}(c)).

At each shake-up satellite, the energy for electron excitation is necessary in addition to the energy for one-electron emission, leading to the energy shift from the quasiparticle peak. Then, we also roughly estimated the satellite peak position of [$\underline{p},\underline{q},\overline{r}$] shake-up satellite within the HF level defined as follows:
\begin{equation}
  \varepsilon^{\rm Satellite (HF)}\equiv\varepsilon^{\rm HF}_p + \varepsilon^{\rm HF}_q + \Delta^{\rm HF}_{q,p} - (\varepsilon^{\rm HF}_r + \Delta^{\rm HF}_{r,pq}) \ , 
\end{equation}
where $\Delta^{\rm HF}_{q,p}=\braket{pq||pq}$ and $\Delta^{\rm HF}_{r,pq}=\braket{qr||qr}+\braket{pr||pr}$.   The estimated shake-up satellite-peak positions within the HF level are also listed in the Table. As shown in the Table, the estimated values by the HF give good agreement with FCI and GFCCSD.


\begin{table}[htb]
\renewcommand{\arraystretch}{1.3}
  \caption{Satellite-peak position in the unit of Hartree for C obtained with the GFCCSD, FCI, and estimated by the HF (see the text).}
  \begin{tabular}{lcccc} \hline\hline
    Main configuration~~~~~~~~~~~~  &  ~CCSDGF~ &~~~FCI~~~~& ~~~HF~~~~\\ \hline
$[\underline{2s}$,$\underline{2s}$,$\overline{2p}]$ & -1.19 & -1.12 & -1.24 \\
$[\underline{2p}$,$\underline{2p}$,$\overline{3p}]$ & -1.25 & -1.24 & -1.24 \\
$[\underline{2s}$,$\underline{2p}$,$\overline{3p}]$ & -1.55 & -1.43 & -1.53 \\
$[\underline{2s}$,$\underline{2p}$,$\overline{3p}]$ & -1.71 & -1.68 & -1.64 \\
$[\underline{2s}$,$\underline{2s}$,$\overline{3p}]$ & -2.05 & -2.03 & -1.93 \\ \hline\hline
  \end{tabular}
  \label{Table2}
\end{table}

The satellite-peak position for C atom is listed in Table~\ref{Table2}. The analysis of the satellite peaks is complicated than those of Ne atom due to the existence of the exchange splitting in $1s$ and $2s$ orbitals. Accordingly, the shake-up satellite has many branches. In Table~\ref{Table2}, many different shake-up satellites are generated from the same electron configuration. This behavior is a fingerprint of the exchange splitting in $1s$ and $2s$ orbitals.

\subsection{One-electron energy spectra from H to Ne}\label{result3}

\begin{figure*}
\includegraphics[width=1.0\linewidth]{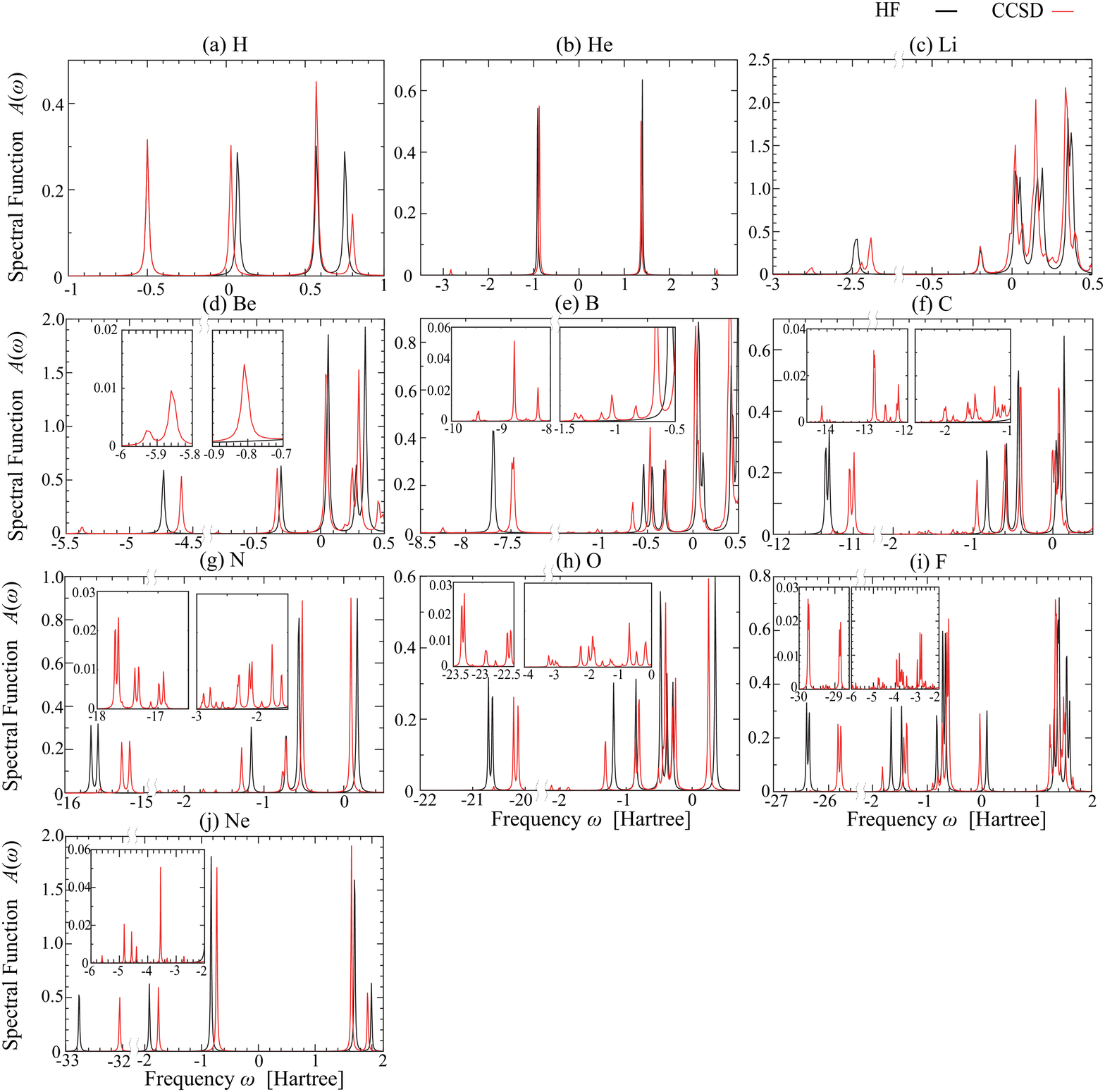}
\caption{(Color online) Single-electron spectrum of H(a), He(b), Li(c), Be(d), B(e), C(f), N(g), O(h), F(i), and Ne(j) obtained by the HF (black line) and the GFCCSD (red line) with the cc-pVDZ basis set. Each inset shows the enlarged satellite peaks. The energy levels of HOMO for each atom are listed in Table.~\ref{HOMO}.}
\label{Fig2}
\end{figure*}

\begin{table}[htb]
  \caption{Energy levels of HOMO in Hartree for each atom calculated by GFCCSD (with cc-pVDZ basis), $\Delta$CCSD (with cc-pVDZ basis), and experiments.}
  \begin{tabular}{lccc} \hline\hline
    atom~~  & GFCCSD & $\Delta$CCSD (previous work) \cite{experiment} &~~Experiment\cite{experiment}~~~\\ \hline 
    H   &   $-0.50$ &  $-0.499$ & $-0.500$ \\
    He  &   $-0.88$ &  $-0.894$ & $-0.904$  \\
    Li  &   $-0.20$ &  $-0.196$ & $-0.198$  \\
    Be  &   $-0.34$ &  $-0.341$ & $-0.343$ \\
    B   &   $-0.30$ &  $-0.296$ & $-0.305$  \\
    C   &   $-0.41$ &  $-0.403$ & $-0.414$  \\ 
    N   &   $-0.52$ &  $-0.522$ & $-0.534$  \\ 
    O   &   $-0.47$ &  $-0.472$ & $-0.500$  \\ 
    F   &   $-0.61$ &  $-0.614$ & $-0.640$  \\ 
    Ne  &   $-0.72$ &  $-0.768$ & $-0.792$  \\ 
\hline\hline
  \end{tabular}
  \label{HOMO}
\end{table}

In this subsection, we show single-electron spectra with a more accurate basis set of the cc-pVDZ set, including 1s, 2s, and 2p orbitals for H and He, and 1s, 2s, 3s, 2p, 3p, and 3d orbitals for Li to Ne. The quasiparticle peaks of closed shell systems such as He and Be atoms show no exchange splitting in deep energy regions like Ne atom described in the subsec.~\ref{result2}.
On the other hand, the open-shell atoms exhibit exchange splitting indicating the breaking of spin symmetry like C atom. For all the atoms, we can confirm that the GFCCSD calculations give smaller HOMO-LUMO gap than the HF. 

It is noteworthy that F atom energetically prefers the negatively charged state in GFCCSD with the cc-pVDZ basis set, i.e. ${\rm F^{-}}$: The energy level of the LUMO of F atom is located below the vacuum level. This gives a good correspondence to the experimental facts \cite{NegAtom}. In contrast, the LUMO level calculated by the HF is positive. The negative ion ${\rm H^{-}}$ and ${\rm O^{-}}$ are also stable experimentally, but our GFCCSD calculations did not show the correct behaviors because of our still poor basis set, namely cc-PVDZ is not accurate enough for describing the delocalized LUMO states. To reproduce the experiments quantitatively, we need more atomic orbitals or have to use the complete basis set in our calculations.

\begin{table}[htb]
  \caption{Quasiparticle peaks of Ne in Hartree based on the HF and the GFCCSD with the cc-pVDZ basis set and the DFT with the complete basis set.}
  \begin{tabular}{lccc} \hline\hline
    Orbital~~~~~~~~~~~~~~~~ & ~~~~~HF~~~~~ & ~~~GFCCSD~~~ & ~~~~DFT~~~~ \\ \hline 
    $1s$    & -32.765 & -32.05 & -30.489  \\
    $2s$    &  -1.919 &  -1.76 &  -1.333  \\
    $2p$    &  -0.832 &  -0.73 &  -0.491  \\ \hline\hline
  \end{tabular}
  \label{Table3}
\end{table}

\begin{table}[htb]
  \caption{Quasiparticle peaks of C in Hartree based on the HF and the GFCCSD with the cc-pVDZ basis set and the DFT with the complete basis set.}
  \begin{tabular}{lccc} \hline\hline
    Orbital~~~~~~~~~~~~~~~~ & ~~~~~HF~~~~~ & ~~~GFCCSD~~~ & ~~~~DFT~~~~ \\ \hline
    $1s$    & -11.345 & -11.05 & -10.071  \\
            & -11.301 & -10.99 &          \\ 
    $2s$    &  -0.826 &  -0.95 &  -0.509  \\
            &  -0.583 &  -0.60 &          \\ 
    $2p$    &  -0.434 &  -0.41 &  -0.197  \\ \hline\hline
  \end{tabular}
  \label{Table4}
\end{table}

Lastly, we compare the quasiparticle peaks of the GFCCSD with those of the DFT. In the DFT calculations, we adopted the spherical symmetry for each atom, which the exact DFT scheme should keep, and solved the Kohn-Sham equations along the radial direction by introducing the real space grids. Table \ref{Table3} shows the quasiparticle peaks of Ne atom. The peaks calculated by the DFT are larger than those of the GFCCSD and the HF since the DFT calculation suffers from the self-interaction error. Table \ref{Table4} shows the quasiparticle peaks of C atom. In this calculation, in order to converge the self-consistent field (SCF) loop in the Kohn-Sham equation, we increased the atomic number by 0.007. Two orbital energies for 1$s$ and 2$s$ are listed, because these orbitals are splitted into two because of the exchange splitting in the HF and GFCCSD as stated in subsec.~\ref{result1}. The exact DFT calculations do not reproduce the splitting. Furthermore, importantly, as explained in detail in subsec.~\ref{result1}, the DFT shows no Mott gap.

We have compared the calculated spectra with experimentally obtainable results for Li atom \cite{Li_experiment}. The quasiparticle peaks with exchange splitting by the GFCCSD method are $-2.44$ and $-2.39$ Hartree. The experimental spectrum also has exchange splitting and their values are $-2.43$ and $-2.37$ Hartree. The satellite peaks of the GFCCSD method and the photoelectron spectroscopy are $-2.76$ and $-2.733$ Hartree, respectively. We have found that the GFCCSD method can reproduce the experimental spectrum including the satellite peaks.

\section{conclusion}\label{conclusion}
In this study, we have calculated single-electron energy spectra via the Green's function based on the coupled-cluster singles and doubles (GFCCSD) method for isolated atoms from H to Ne. In order to check the accuracy of the GFCCSD method, we compared the results with the exact ones calculated from the full-configuration interaction (FCI). Consequently, we have found that the GFCCSD method reproduces not only the correct quasiparticle peaks but also satellite ones by comparing the exact spectra with the 6-31G basis set. It is also found that open-shell atoms such as C atom exhibit Mott gaps at the Fermi level, which the exact density-functional theory (DFT) fails to describe. The GFCCSD successfully reproduces the Mott HOMO-LUMO (highest-occupied molecular orbital and lowest-unoccupied molecular orbital) gaps even quantitatively. We also discussed the origin of satellite peaks as shake-up effects by checking the wave functions responsible for satellite peaks. We have also presented the single-electron energy spectra using a higher accurate basis set, cc-pVDZ and showed that the results give good agreement with experimental results available. The GFCCSD is a novel cutting edge to investigate the electronic states in detail.

\begin{acknowledgments}
This research was supported by MEXT as ``Exploratory Challenge on Post-K computer" (Frontiers of Basic Science: Challenging the Limits). This research used computational resources of the K computer provided by the RIKEN Advanced Institute for Computational Science through the HPCI System Research project (Project ID: hp170261). Y.M. acknowledges the support from JSPS Grant-in-Aid for Young Scientists (B) (Grant No. 16K18075).
\end{acknowledgments}

\bibliographystyle{apsrev4-1}
\bibliography{paper}

\onecolumngrid

\end{document}